\documentclass[twocolumn,prb]{revtex4-1}

\usepackage{graphicx}
\usepackage{txfonts}
\usepackage{epsfig}

\def\avg#1{\left\langle#1\right\rangle}
\def\bra#1{\left\langle#1\right|}
\def\ket#1{\left|#1\right\rangle}

\def\abs#1{\left|#1\right|}
\def\kc#1{\left(#1\right)}

\def\be{\begin{equation}}       \def\ee{\end{equation}}
\def\bea{\begin{eqnarray}}      \def\eea{\end{eqnarray}}
\def\ba{\begin{array} }
\def\ea{\end{array} }
\def\bnum{\begin{enumerate} }
\def\enum{\end{enumerate}}

\def\=>{\Rightarrow}
\def\>{\rightarrow}

\begin{document}
\title{Momentum polarization: an entanglement measure of topological spin and chiral central charge}

\author{Hong-Hao Tu$^1$, Yi Zhang$^2$ and Xiao-Liang Qi$^2$}
\affiliation{$^1$ Max-Planck-Institut f\"ur Quantenoptik, Hans-Kopfermann-Str. 1, D-85748 Garching, Germany\\
$^2$ Department of Physics, Stanford University, Stanford, California 94305, USA}
\date{\today}
\begin{abstract}
Topologically ordered states are quantum states of matter with topological ground state degeneracy and quasi-particles carrying fractional quantum numbers and fractional statistics. The topological spin $\theta_a=2\pi h_a$ is an important property of a topological quasi-particle, which is the Berry phase obtained in the adiabatic self-rotation of the quasi-particle by $2\pi$. For chiral topological states with robust chiral edge states, another fundamental topological property is the edge state chiral central charge $c$. In this paper we propose a new approach to compute the topological spin and chiral central charge in lattice models by defining a new quantity named as the momentum polarization. Momentum polarization is defined on the cylinder geometry as a universal subleading term in the average value of a ``partial translation operator". We show that the momentum polarization is a quantum entanglement property which can be computed from the reduced density matrix, and our analytic derivation based on edge conformal field theory shows that the momentum polarization measures the combination $h_a-\frac{c}{24}$ of topological spin and central charge. Numerical results are obtained for two example systems, the non-Abelian phase of the honeycomb lattice Kitaev model, and the $\nu=1/2$ Laughlin state of a fractional Chern insulator described by a variational Monte Carlo wavefunction. The numerical results verifies the analytic formula with high accuracy, and further suggests that this result remains robust even when the edge states cannot be described by a conformal field theory. Our result provides a new efficient approach to characterize and identify topological states of matter from finite size numerics.
\end{abstract}
\maketitle

\section{Introduction}

Since the discovery of integer and fractional quantum Hall states in 1980's\cite{klitzing1980,tsui1982}, the research on topological states of matter has attracted tremendous experimental and theoretical interest. The recent discovery of topological insulators and topological superconductors have greatly expanded the knowledge of topological states.\cite{hasan2010,qi2011rmp,moore2010} In general, topological states are gapped zero-temperature states of matter which cannot be deformed to trivial states without experiencing a quantum phase transition. Here the ``trivial states" are reference states which have no quantum entanglement between any two different regions of space. Topological states of matter are characterized by topological properties such as gapless edge/surface states, ground state degeneracy, bulk quasi-particle excitations with fractional charge and fractional statistics. The fractionalization of quantum numbers and statistics can only occur in topological states with ground state degeneracy, which are called topologically ordered states\cite{wen1989,wen1990b}.

In a topologically ordered state, a given configuration of topological quasi-particles correspond to a finite dimensional Hilbert space. Braiding quasi-particles around each other leads to a unitary transformation in the Hilbert space, which is named as the fractional statistics of the particles\cite{nayak2008}. When the braiding of different particles  leading to non-commuting transformations, the statistics of the particle is non-Abelian, and otherwise it is Abelian. Besides the statistics, each quasi-particle $a$ also has a fractionalized spin $\theta_a$ which is a phase factor obtained by the state of the system during the self-rotation of $a$ by $2\pi$.  The spins of topological quasi-particles are related to their statistics, because when two particles wind around each other by $2\pi$ and return to their positions, from far away, other particles cannot determine whether there are two particles braiding or there is only one particle (obtained by the ``fusion" of the two particles) spinning around itself.\cite{zhwang2010} The spin-statistics relation is a generalization of that for ordinary particles--{\it i.e.}, fermions have half-odd-integer spin while bosons have integer spin. Therefore the spin values of quasi-particles are an important set of information that distinguishes different topological states. The calculation of topological spin is generically difficult except for algebraically defined topological field theories\cite{zhwang2010} and some ideal model states\cite{wen2008,bernevig2009,read2009}. For microscopic lattice models with a topological phase, the topological spin of quasi-particles has only been computed numerically for several special cases: 1) The honeycomb lattice Kitaev model studied by calculating the braiding of quasi-particles\cite{lahtinen2009,bolukbasi2012}; 2) The fractional Chern insulators studied by modular transformations on a torus\cite{zhang2012a,zhang2012b,cincio2012}. Such numerical results are usually restricted by small system size and specific knowledge about the ground state wavefunction.

For chiral topological states with robust chiral edge states, another fundamental topological quantity is the edge state chiral central charge. If the edge state is described by a chiral conformal field theory (CFT), it has a chiral central charge $c$, which determines the heat current $I_E=\frac{\pi}{6}cT^2$ at a given temperature.\cite{affleck1986} The central charge also appears in the gravitational anomaly of the edge\cite{alvarezgaume1984} if the system is coupled to a gravitational field. Since the edge states of a topological state is only a well-defined one-dimensional state in energy below the bulk gap, it can at most be described by a CFT in the long wavelength limit. (As we will discuss later, there are examples where the edge state is not Lorentz invariant even in the long wavelength limit, so that it's not a CFT at all.)

\begin{figure}[tb]
\centerline{
\includegraphics[width=3.5in]{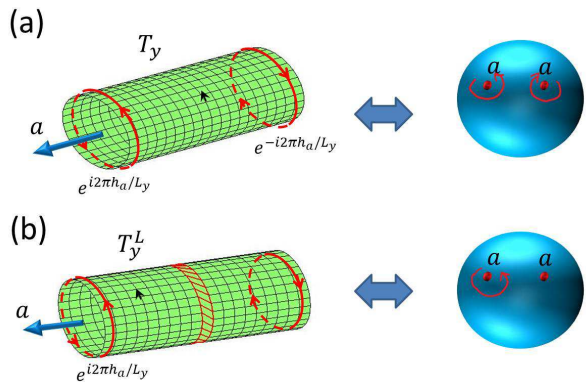}
}
\caption{
\label{fig1}
(a) A global translation of the cylinder is equivalent to spinning two quasiparticles on a sphere to opposite directions. (b) The partial translation $T_y^L$ of the left part of the cylinder. In both panels, the red circles with arrows indicate the chiral edge states.
}
\end{figure}

In this paper, we propose a new numerical method for the calculation of topological spins and chiral central charge in candidate systems of topological states, which is significantly simpler than previous numerical methods, and can be generally applied to identify topological states obtained in numerical approaches such as exact diagonalization and Monte Carlo. This method applies to a topological state defined on cylinder. Since a cylinder is topologically equivalent to a sphere with two punctures, each topological quasiparticle $a$ corresponds to a ground state on the cylinder, which has quasiparticle $a$ in each puncture. This is illustrated in Fig. \ref{fig1} (a). A rotation of the cylinder along the periodic direction $y$ is equivalent to rotating the two particles simultaneously. By mapping to sphere we see that with reference to the direction normal to the sphere, one particle is rotated clockwisely while the other is rotated counter-clockwisely. Consequently, when the particles have spin $\theta_a$, a $2\pi$ rotation of the cylinder should lead to a Berry phase of $e^{i\theta_a}$ from one end and $e^{-i\theta_a}$ from the other end, which cancels each other as expected, since this global rotation is a symmetry of the system. To avoid the cancelation, we need to define a ``twist" of the cylinder where only the left half of the cylinder is rotated, as is illustrated in Fig. \ref{fig1} (b). For a lattice system, we do not have a way to twist the cylinder continuously, but have to twist discretely. If we consider the twist  Fig. \ref{fig1} (b) where half of the cylinder is twisted by one lattice constant, and there are $L_y$ lattice sites in the $y$ direction, such a twist should lead to a Berry's phase of $e^{i\theta_a/L_y}$ coming from the left edge of the cylinder with particle type $a$, without cancelation from the right edge. However, the twist looks like a violent operation since it induces a discontinuous jump at the boundary between left and right halves of the cylinder. Naively this appeared to induce a non-universal correction to the phase obtained during the twist, and thus make it impossible to measure the spin $\theta_a$. The key result of the current work is that the non-universal contribution is independent from topological sector $a$, which is shown in the following formula:

\bea
\lambda_a\equiv \bra{G_a}T_y^L\ket{G_a}&=&\exp\left[\frac{2\pi i}{L_y}p_a-\alpha L_y\right]\label{result}\\
p_a&\equiv &h_a-\frac{c}{24}\nonumber
\eea
with $\ket{G_a}$ the ground state in the $a$ sector, $c$ is the central charge of the edge CFT, and $\alpha$ is a non-universal complex constant. The key property of this formula is that the first term is universal and contains the topological information $h_a$ and $c$, while the second non-universal term is independent from $a$. We name the coefficient $p_a$ of the first term as the {\bf momentum polarization}, which has the physical meaning of the net $y$ momentum carried by the left edge in the ground state (in unit of $\frac{2\pi}{L_y}$). We choose the unit $\frac{2\pi}{L_y}$ which is the momentum quantum carried by a local bosonic excitation. Therefore $p_a\in \mathbb{Z}$ corresponds to a local excitation and the fractional part $p_a$ mod $1$ is the topological contribution determined by the ground state. As we will show below, this result can be understood based on entanglement properties of topological states.

The rest of the paper is organized as following. In Sec. \ref{sec:general} we provide an analytic derivation of  Eq. (\ref{result}) based on edge state conformal field theory and quantum entanglement properties of the topological state. In Sec. \ref{sec:kitaev} we verify this idea by numerical calculations in the honeycomb lattice Kitaev model\cite{kitaev2006}. In the non-Abelian phase of Kitaev model, we find that the spin of the non-Abelian anyon $\sigma$ is $\theta_a=\frac{\pi}8$, in consistency with the expectation that this state is described by an Ising topological quantum field theory.\cite{kitaev2006} In Sec. \ref{sec:fci} this result is further verified in another topological state, the fractional Chern insulators\cite{tang2011,sun2011,sheng2011,neupert2011}. The variational Monte Carlo wavefunction\cite{zhang2012a} of lattice Laughlin $1/2$ state is confirmed to have central charge $c=1$ and quasiparticle charge $\theta_a=\frac{\pi}2$. Finally Sec. \ref{sec:conclusion} is the conclusion and discussions.

\section{Analytic result}\label{sec:general}

In this section we will derive Eq. (\ref{result}) based on the assumption that the edge state of the topological state is a chiral conformal field theory. As will be seen later in Sec. \ref{sec:kitaev}, the numerical results indicate that the applicability of formula (\ref{result}) goes {\it beyond} the edge CFT assumption,  but analytically the result can only be derived under this assumption.

To begin with, we notice that the partial translation $T_y^L$ only acts on the left half of the system, so that we can trace over the right half and define the reduced density matrix $\rho_{La}={\rm tr}_R\left(\ket{G_a}\bra{G_a}\right)$ such that
\bea
\lambda_a\equiv \bra{G_a}T_y^L\ket{G_a}={\rm tr}_L\left(\rho_{La}T_y^L\right)\label{lambda}
\eea
The reduced density matrix $\rho_{La}$ is generically difficult to compute, but for topologically ordered states with chiral edge states, some long wavelength behavior of $\rho_{La}$ is known. In 2008, H. Li and F. D. M. Haldane observed that the entanglement spectrum ({\it. i.e.} the spectrum of the "entanglement Hamiltonian" $H_E$ defined as $H_E=-\log\rho_{La}$ ) is qualitatively the same as that of the edge Hamiltonian\cite{li2008}. This is proposed as a generic feature of the topological states, which has been verified numerically and analytically in various physical systems such as general fractional quantum Hall wavefunctions\cite{thomale2010,chandran2011}, free fermion topological states\cite{fidkowski2010,turner2010} and Kitaev model\cite{yao2010}, etc. A general analytic derivation of the entanglement-edge state correspondence has been proposed in Ref. \cite{qi2012} for a generic topological state with chiral CFT edge states. Here we will briefly review the result of Ref. \cite{qi2012} which will be applied to the current work. 

\begin{figure}[tbp]
\centerline{
\includegraphics[width=3.5in]{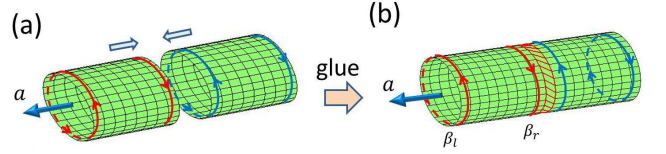}
}
\caption{
\label{fig2}
(a) A cylinder can be considered as two separate cylinders glued together. Each cylinder have chiral edge states (red and blue circles) and the gluing of the two cylinders couples the edge states at the interface. (b) After gluing the two parts of the cylinder are entangled through the interface, and the reduced density matrix of the left half cylinder is described by a thermal density matrix of the edge state CFT with finite ``entanglement temperature" $\beta_r^{-1}$ at the right edge, and zero entanglement temperature $\beta_l^{-1}=0$ at the left edge (see text).
}
\end{figure}

The basic idea behind this derivation is that a cylinder can be considered as two cylinders $A, B$ glued together, as is illustrated in Fig. \ref{fig2} (a). Before the gluing procedure, each cylinder has two edge states on the two boundaries, which in the long wavelength limit are described by a conformal field theory Hamiltonian $H_{A(B)}=H_{A(B)l}+H_{A(B)r}$, with $l,r$ denoting the left and right edge of each cylinder. A topological sector of the cylinder labeled by quasiparticle $a$ corresponds to conformal blocks of the edge CFT formed by a primary field with scaling dimension $h_a$ and its descendants. The ``gluing" is done by turning on a coupling between the two cylinders across the edge, and the coupling can be considered as a relevant coupling in the edge CFT. Using the property of CFT with relevant coupling, Ref. \cite{qi2012} reaches the conclusion that the reduced density matrix of the edge state $Ar$ is $\rho_{Ar}^{(a)}=Z_{ra}^{-1}\left.e^{-\beta_rH_{Ar}}\right|_a$, which is a thermal density matrix of the chiral CFT restricted to the sector $a$. The left edge of $A$ region is far from the $B$ region, so that the left edge CFT stays in its ground state, with no entanglement with $B$ region. This can be described by the same thermal density matrix with a temperature $\beta_l\rightarrow \infty$. The total reduced density matrix of the left half cylinder ($A$ region) is
\bea
\rho_{La}=\rho_{Al}^{(a)}\otimes \rho_{Ar}^{(a)}=Z^{-1}_a\left.e^{-\beta_lH_{Al}-\beta_rH_{Ar}}\right|_a, ~\beta_l=\infty,~\beta_r\text{~finite}\label{dmCFT}
\eea
as is illustrated in Fig. \ref{fig2} (b).
Therefore $\lambda_a$ defined in Eq. (\ref{lambda}) is the average value of translation operator $T_y^L$ in a CFT with left edge at zero temperature, and right edge at finite temperature. In CFT the translation operator has the form
\bea
T_y^L=e^{i(P_l+P_r)\frac{2\pi}{L_y}}=e^{i(H_l-H_r)\frac{2\pi}{vL_y}}\label{TyCFT}
\eea
Here we have used the linear dispersion of the edge CFT $H_l=vP_l,~H_r=-vP_r$ with $v$ the velocity of the CFT. Now we introduce the characters $\chi_a(q)$ of the CFT\cite{francesco1996}$^,$\footnote{For a short and concise review of some basic properties about CFT used here, {\it c. f. } Ref. \onlinecite{cardy2004}} which are defined as
\bea
\chi_a(q)={\rm tr}_a(q^{L_0})={\rm tr}_a(q^{\frac{L_y}{2\pi v}H_l})={\rm tr}_a(q^{\frac{L_y}{2\pi v}H_r})
\eea
with ${\rm tr}_a$ the trace in the conformal block $a$. From Eq. (\ref{dmCFT}) and (\ref{TyCFT}), $\lambda_a$ can be expressed in the character $\chi_a$ by
\bea
\lambda_a=\frac{\chi_a\kc{e^{ \frac{2\pi}{L_y}\left(i-v\beta_l\right)}}\chi_a\kc{e^{ \frac{2\pi }{L_y}\left(-i-v\beta_r\right)}}}{\chi_a\kc{e^{-\frac{2\pi}{L_y} v\beta_l}}\chi_a\kc{e^{-\frac{2\pi}{L_y} v\beta_r}}}\label{character}
\eea

Now we consider a large but finite $L_y$, such that $v\beta_r\ll L_y\ll v\beta_l=\infty$. The left edge is in zero temperature, so that the character is dominated by the ground state contribution $\chi_a(q)=q^{h_a-\frac{c}{24}}$. Consequently ${\chi_a\kc{e^{ \frac{2\pi}{L_y}\left(i-v\beta_l\right)}}}/{\chi_a\kc{e^{-\frac{2\pi}{L_y} v\beta_l}}}=\exp\left[i\frac{2\pi}{L_y}\left(h_a-\frac{c}{24}\right)\right]$.  On the other hand, the right edge is in the high temperature limit since $v\beta_r\ll L_y$. To compute the character in that limit we can make use of the modular transformation property of the character
\bea
\chi_a\left(e^{-2\pi i/\tau}\right)=\sum_bS_{ab}\chi_b(e^{i2\pi \tau})
\eea
with $S_{ab}$ the modular $S$ matrix\cite{francesco1996}. Taking $e^{-2\pi i/\tau}=e^{-\frac{2\pi}{L_y}v\beta_r}$ we have
\bea
\chi_a(e^{-\frac{2\pi}{L_y}v\beta_r})=\sum_bS_{ab}\chi_b\left(e^{-\frac{2\pi L_y}{v\beta_r}}\right)\label{modular}
\eea
Since the right-hand side is in low temperature limit, we can approximate $\chi_b\left(e^{-\frac{2\pi L_y}{v\beta_r}}\right)\simeq e^{-\frac{2\pi L_y}{v\beta_r}\left(h_b-\frac{c}{24}\right)}$. In the low temperature limit the trivial sector $\chi_1$ with $h_1=0$ dominates all other sectors with positive $h_a$, so that we have $\chi_a(e^{-\frac{2\pi}{L_y}v\beta_r})\simeq S_{a1} e^{\frac{2\pi L_y}{v\beta_r}\frac{c}{24}}$. The same modular transformation can be applied to the numerator in Eq. (\ref{character}) and leads to $\chi_a\kc{e^{ \frac{2\pi }{L_y}\left(-i-v\beta_r\right)}}\simeq S_{a1}e^{\frac{2\pi L_y}{i+v\beta_r}\frac c{24}}$. In summary we obtain
\bea
\lambda_a=\exp\left[i\frac{2\pi}{L_y}\left(h_a-\frac{c}{24}\right)-L_y\frac{2\pi c}{24}\frac{i}{v\beta_r\left(i+v\beta_r\right)}\right]
\eea
in the limit of $v\beta_r\ll L_y$, which demonstrates Eq. (\ref{result}) with $\alpha=\frac{2\pi c}{24}\frac{i}{v\beta_r\left(i+v\beta_r\right)}$ non-universal, but independent from topological sector $a$. It should be noted that ${\rm Re}\alpha= \frac{2\pi c}{24}\frac{1}{v\beta_r\left(1+v^2\beta_r^2\right)}>0$, which means $\lambda_a$ decays exponentially when the size $L_y$ increases.

For finite $L_y/v\beta_r$, the formula above obtains finite size corrections coming from two origins: i) the contribution of excited states to $\chi_a(q)$ in low temperature limit. $\chi_a(q)=q^{h_a-\frac{c}{24}}\left(1+n_{1a}q+...\right)$ with $n_{1a}$ the number of states in the topological sector $a$ with momentum $2\pi/L_y$. ii) The contribution of other topological sectors $b$ in the expansion (\ref{modular}), which is proportional to $q^{h_b}$. In the two corrections above, the most leading contribution comes from the ground state of the topological sector with minimal scaling dimension. Denoting the minimal scaling dimension by $h_{\rm min}$, we obtain
\bea
\chi_a\left(e^{-\frac{2\pi}{L_y}v\beta_r}\right)\simeq S_{a1} e^{\frac{2\pi L_y}{v\beta_r}\frac{c}{24}}\left[1+O\left(e^{-\frac{2\pi L_y}{v\beta_r}h_{\rm min}}\right)\right]
\eea
and similarly for the numerator of Eq. (\ref{character}). Consequently the finite size correction to $\lambda_a$ is
\bea
\lambda_a&=&\exp\left[i\frac{2\pi}{L_y}\left(h_a-\frac{c}{24}\right)-\alpha L_y\right]\left(1+O\left(e^{-{\rm Re}(\alpha)L_y}\right)\right)\label{errorbar}\\
\text{with~}\alpha&=&\frac{2\pi c}{24}\frac{i}{v\beta_r\left(i+v\beta_r\right)}\nonumber
\eea
We see that the finite size correction to formula (\ref{result}) decays exponentially versus $L_y/v\beta_r$, which suggests that this method can be applied well to finite size numerical calculations. Such an exponential convergence behavior is confirmed in the numerical results presented in next section.

Besides the finite size correction, other corrections to the formula (\ref{result}) may occur in a realistic system due to deviations of the reduced density matrix $\rho_{La}$ from the CFT behavior. For example the edge state dispersion is not strictly linear, and the state counting may deviate from that of the pure CFT at high energy. Somewhat surprisingly, in the two physical systems we studied numerically, such nonuniversal corrections have not been found, and the result (\ref{result}) remains valid even if the edge state deviates strongly from a CFT.

\section{Numerical results on honeycomb-lattice Kitaev model}\label{sec:kitaev}

\begin{figure}[tb]
\centerline{
\includegraphics[width=3in]{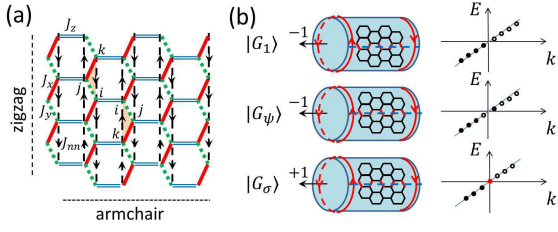}
}
\caption{
\label{fig3}
(a) Illustration of the definition of honeycomb lattice Kitaev model in Eq. (\ref{HKitaev}). The $x,y,z$ links are shown by the red solid lines, green dash lines and blue double lines, respectively. The second neighbor couplings are defined on the triangles shaded in orange color, with the labeling $i,j,k$ of the triangle shown on the figure. (b) Illustration of the three ground states in the three topological sectors of the non-Abelian phase of Kitaev model. The right panels show the edge state energy spectrum with solid and hollow circles stand for occupied and unoccupied quasi-particle states. $\ket{G_1}$ and $\ket{G_\psi}$ correspond to the ground state and the lowest energy quasiparticle excitation state with fermion anti-periodic boundary conditions, while $\ket{G_\sigma}$ correspond to one of the degenerate ground states with fermion periodic boundary condition. The red dot in the right panel marks the Majorana zero mode, the occupation of which does not change the momentum polarization of $\ket{G_\sigma}$.
}
\end{figure}

In this section we will calculate $\lambda_a$ numerically for the honeycomb-lattice Kitaev model\cite{kitaev2006}. Kitaev model is a special spin $1/2$ model which has a non-Abelian topological phase that we are interested in. For our purpose, we consider the following Hamiltonian of the Kitaev model with nearest neighbor and second neighbor interactions:
\bea
H&=&\sum_{\avg{ij}\in x\text{~links}}J_x \sigma_{ix}\sigma_{jx}+\sum_{\avg{ij}\in y\text{~links}}J_y \sigma_{iy}\sigma_{jy}+\sum_{\avg{ij}\in z\text{~links}}J_z \sigma_{iz}\sigma_{jz}\nonumber\\
& &+J_{nn}\sum_{(ijk)\in \triangle}\sigma_{iy}\sigma_{jz}\sigma_{kx}\label{HKitaev}
\eea
with $x,y,z$ links the three types of nearest neighbor links on the honeycomb lattice shown in Fig. \ref{fig3} (a). $\triangle$ in the $J_{nn}$ term denotes the upper and lower triangles shown in Fig. \ref{fig3} (a), in which $ijk$ runs counter-clockwisely around the triangle and $j$ is at the upper or lower corner. For simplicity, we only introduced one type of second neighbor coupling instead of including all possible terms obtained by $60$ degree rotations.

The unique property of the Kitaev model is that it can be solved by the following Majorana representation
\bea
\sigma^a_i=i\gamma^a_i\eta_i
\eea
in which $\gamma_i^a, a=x,y,z$ and $\eta_i$ are four Majorana fermion operators defined at site $i$. The Hilbert space of 4 Majorana fermions is 4-dimensional. The two physical states of the spin at site $i$ form a subspace of the Majorana fermion Hilbert space which is determined by the condition
\bea
D_i\equiv i\gamma^x_i\gamma^y_i\gamma^z_i\eta_i=-i\sigma_i^x\sigma_i^y\sigma_i^z=1\label{constraint}
\eea
Using this representation the Hamiltonian is transformed to
\bea
H&=&-\sum_{a=x,y,z}\sum_{\avg{ij}\in a \text{links}}J_au_{ij}i\eta_i\eta_j-J_{nn}\sum_{\avg{ik}\in \text{nn links}}u_{ik}i\eta_i\eta_k\nonumber\\
\text{with~}u_{ij}&=&i\gamma_i^a\gamma_j^a,\text{for~}\avg{ij}\in a\text{~links}\nonumber\\
u_{ik}&=&u_{ij}u_{jk},\text{for~}\avg{ik}\in\text{nn links},~(ijk)\in \triangle
\eea
Since $\left[u_{ij},u_{kl}\right]=0$ for all different links $\avg{ij},\avg{kl}$, the Hamiltonian can be viewed as a free Majorana fermion Hamiltonian of $\eta_i$ with $u_{ij}=\pm 1$ taking its eigenvalues. However the physical states have to satisfy the constraint (\ref{constraint}), which can be obtained by a projection in the following form:
\bea
\ket{\Psi}=\prod_i\frac{D_i+1}2\ket{\Psi_F(\left\{u\right\})}\otimes \ket{\left\{u\right\}}
\eea
with $D_i$ defined in Eq. (\ref{constraint}). The link variable $u_{ij}$ can be viewed as a $Z_2$ gauge field coupled to the Majorana fermion, and $D_i$ acts as a $Z_2$ gauge transformation which transforms $\eta_i\rightarrow -\eta_i$ and $u_{ij}\rightarrow -u_{ij}$ for all links connected to $i$. The physical state $\ket{\Psi}$ is a superposition of all gauge equivalent configurations of gauge field $u_{ij}$ in direct product with the corresponding fermion state $\ket{\Psi_F(\left\{u\right\})}$.
Kitaev model has several phases including a gapless phase and Abelian and non-Abelian topologically ordered phases.\cite{kitaev2006} In the following we mainly discuss the non-Abelian phase in which the $\eta_i$ fermion has a band structure with a nontrivial Chern number $C=1$. There are three types of topological quasiparticles $1,\sigma,\psi$ in which $1$ is the vacuum, $\psi$ is the fermionic excitation of $\eta$ Majorana fermion, and $\sigma$ is a $Z_2$ flux of the gauge field $u_{ij}$, with a Majorana zero mode of $\eta$ Majorana fermion trapped in it. $\sigma$ is the non-Abelian quasi-particle. Corresponding to the three quasi-particle sectors, there are three topological sectors for the system on a cylinder. Depending on the flux of $u_{ij}$ gauge field around the cylinder, the fermion $\eta_i$ has periodic or anti-periodic boundary conditions. As is shown in Fig. \ref{fig3} (b), the Majorana fermion has chiral edge states due to the topological band structure. When the $Z_2$ flux in the cylinder is $-1$, the edge state has anti-periodic boundary condition and momentum eigenvalues $k=\frac{2\pi}{L_y}\left(n+\frac 12\right)$. Denote the $\eta$ fermion ground state in this sector as $\left|G_F^-\right\rangle$, we have $\ket{G_1}=P\ket{G_F^-}\otimes \ket{\left\{u_0^-\right\}}$ with $P=\prod_i(D_i+1)/2$ the projection to gauge invariant states. $u_0^-$ is a gauge configuration with flux $-1$ in the cylinder. For example the gauge choice in Fig. \ref{fig3} (b) can be made, with $u_{ij}=-1$ for all red links across the dashed line, and $u_{ij}=1$ otherwise. The $\psi$ sector is the lowest energy excitation state $\ket{G_\psi}=Pf^\dagger_{Lk}f^\dagger_{R-k}\ket{G_F^-}\otimes \ket{\left\{u_0^-\right\}}$, with $f^\dagger_{L,Rk}$ the creation operator of the left and right edge states, respectively, and $k=\frac{\pi}{L_y}$. The third sector $\sigma$ corresponds to the periodic boundary condition $\ket{G_\sigma}=P\ket{G_F^+}\otimes \ket{\left\{u_0^+\right\}}$. $u_0^+$ is a guage configuration with $+1$ flux in the cylinder, which can be taken as $u_{ij}=1$ for all $\avg{ij}$. It should be noticed that there are two degenerate states before projection due to the Majorana zero modes, but only one survives in the projection.

Due to the absence of quantum fluctuation of the $Z_2$ gauge field, the entanglement entropy and entanglement spectrum of Kitaev model can be obtained rigorously\cite{yao2010}. Using the reduced density matrix obtained in Ref. \cite{yao2010}, $\lambda_a$ in Eq. (\ref{lambda}) can be reduced to that of the free Majorana fermion: $\lambda_a={\rm tr}\left(\rho_{Fa}T_y^{LF}\right)$. Here $\rho_{Fa}$ is the reduced density matrix of the left half cylinder, and $T_y^{LF}$ is the gauge covariant translation operator acting on the fermion $\eta_i$, which does a translation of the $\eta_i$ fermions associated with a gauge transformation that translates the gauge field configuration $u_{ij}$. This is like the magnetic translation operators in a Landau level. More details on the reduced density matrix and the calculation of momentum polarization are given in Appendix \ref{app:kitaev}. As a general result for free fermions and free bosons\cite{peschel2003}, $\rho_{Fa}$ can be determined by the two-point function $\avg{\eta_i\eta_j}$ (due to the Wick theorem), and the entanglement Hamiltonian has the quadratic form $H_E=\log Z_a-\log \rho_{Fa}=\sum_{i,j\in L}iA_{ij}\eta_i\eta_j$. The constant $Z_a$ is determined by the normalization condition of the reduced density matrix. On the half cylinder, the entanglement Hamiltonian can be diagonalized into the form
\bea
H_E=\sum_{n,k}\xi_{nk}\left(f_{nk}^\dagger f_{nk}-\frac 12\right)\label{HEfree}
\eea
with $k$ the $y$-direction momentum taking the values of $\frac{2\pi}{L_y}n$ and $\frac{2\pi}{L_y}\left(n+\frac12\right)$ ($n=0,1,2,...,L_y-1$) for periodic and anti-periodic boundary conditions, respectively. $f_{nk}$ are the quasiparticle annihilation operators and $\xi_{nk}$ are the eigenvalues of the entanglement Hamiltonian. In the momentum basis, the translation operator simply multiples a phase $e^{ik}$ to each quasi-particle operator $f_{nk}$, so that
\bea
T_y^{LF}=\exp\left[i\sum_{n,k}kf_{nk}^\dagger f_{nk}\right]\label{Tyfree}
\eea
Using Eq. (\ref{HEfree}) and Eq. (\ref{Tyfree}) we obtain
\bea
\lambda_a&=&Z^{-1}_a{\rm tr}\left(e^{-H_E}T_y^{LF}\right)\nonumber\\
&=&\prod_{n,k}\left[\frac{1+e^{ik}}2+\frac{1-e^{ik}}2\tanh\frac{\xi_{nk}}2\right]\label{lambdafree}
\eea
Here $a=1,\sigma$ corresponds to the anti-periodic and periodic boundary conditions, respectively. Since the $\psi$ state is an excited state of the anti-periodic sector, it is not directly included in the formula above. However, it is straightforward to show that $\lambda_\psi=\lambda_1e^{i\pi/L_y}$ with the phase factor $e^{i\pi/L_y}$ contributed by the edge state fermion excitation $f_{Lk}^\dagger$.

\begin{figure}[tb]
\centerline{
\includegraphics[width=3.5in]{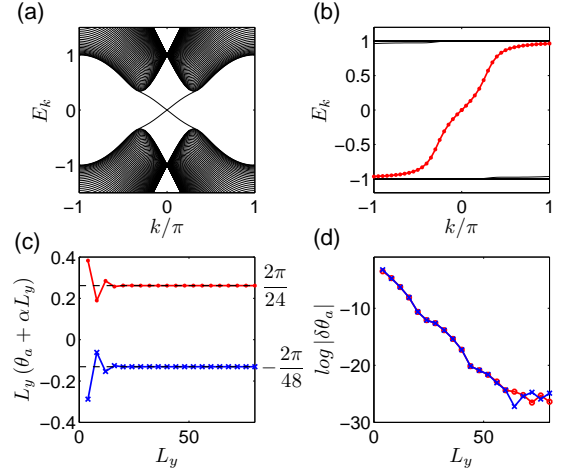}
}
\caption{(a) The energy spectrum and (b) the entanglement spectrum ($\xi_{nk}$ in Eq. (\ref{HEfree})) of the Kitaev model.
(c) The value of $L_y\left(\theta_a(L_y)+\alpha L_y\right)$ versus the system size $L_y$ for $a=1$ (blue $\times$) and $\sigma$ (red $\circ$) sectors. $\theta_a={\rm Im}\log \lambda_a$. $\alpha$ is obtained from the slope of a linear fitting of $L_y\theta_a(L_y)$ versus $L_y^2$. The dashed lines mark the theoretical values given by Eq. (\ref{result}) $2\pi\left(h_a-\frac{c}{24}\right)$ for $h_1=0,~h_\sigma=\frac1{16}$. (d) The deviation of $\theta_a$ from the theoretical value $\delta\theta_a=\theta_a+\alpha L_y-\frac{2\pi}{L_y}\left(h_a-\frac{c}{24}\right)$ versus $L_y$ plotted in a log scale. The calculations are done for a system with $L_x=40$ along the $x$ direction with zigzag edge and the parameters $J_x=J_y=J_z=1,~J_{nn}=0.2$.
\label{fig4}}
\end{figure}

 A honeycomb lattice on the cylinder can be defined with different orientations. Two different simple orientations are known as the ``armchair edge" and "zigzag edge" in the graphene literature, as are illustrated in Fig. \ref{fig3} (a). We will focus on the zigzag edge, which is convenient for the reason that will be clear in later part of this section. The numerical results of $\lambda_\sigma$ and $\lambda_1$ are shown in Fig. \ref{fig4}. To compare with the formula (\ref{result}), we define $\theta_a(L_y)={\rm Im}\log \lambda_a$ and do a linear fitting of $L_y\theta_a$ versus $L_y^2$. Eq. (\ref{result}) predicts $L_y\theta_a(L_y)=- {\rm Im}\alpha L_y^2+2\pi p_a$, so that the intercept of this fitting at $L_y=0$ gives the numerical value of $h_a$ and $c$. Fig. \ref{fig4} (a) shows $2\pi p_a=L_y\theta_a(L_y)+{\rm Im}\alpha L_y^2$
with the slope ${\rm Im}\alpha$ obtained from linear fitting. The agreement with formula (\ref{result}) is apparent. As is shown in Fig. \ref{fig4} (b), the deviation $\delta\theta_a=\theta_a(L_y)-\left[-{\rm Im}\alpha L_y+\frac{2\pi}{L_y}p_a\right]$ from the theoretical value decays exponentially versus $L_y$, in agreement with the error estimation given by the CFT in Eq. (\ref{errorbar}).

\begin{figure}[tb]
\centerline{
\includegraphics[width=3.5in]{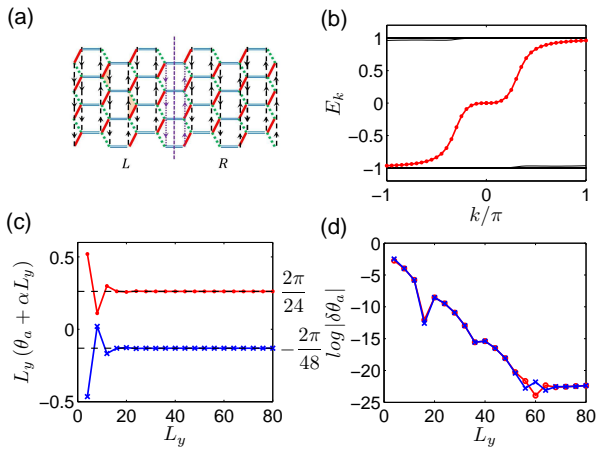}
}
\caption{(a) The Hamiltonian of the Kitaev model is modified by turning off the second neighbor hoppings (purple dotted lines with arrows) along the boundary between left and right parts of the cylinder (purple dashed line). (b) The entanglement spectrum for this geometry, with quartic dispersing entanglement edge states. (c) The value of $L_y\left(\theta_a(L_y)+\alpha L_y\right)$ and (d) the deviation of $\theta_a$ from the theoretical value for the modified geometry. The parameters are the same as in Fig. \ref{fig4}.
\label{fig5}}
\end{figure}

An important question is how robust is the formula (\ref{result}). Since the derivation given in Sec. (\ref{sec:general}) is based on the CFT description of the edge states, it is not obvious whether the deviation of the edge state from CFT behavior will lead to corrections to the universal $1/L_y$ term in Eq. (\ref{result}). Interestingly, the numerical result above shows that the formula applies to the lattice model without any correction to the $1/L_y$ term. To probe the robustness of the formula (\ref{result}), we can modify the Hamiltonian near the cut between left and right parts of the cylinder. As is shown in Fig. \ref{fig5} (a), we turn off the second neighbor hopping along the sites that are at the boundary between left and right parts. Such an edge term does not change the topology of the system, but changes the entanglement spectrum. As is shown in Fig. \ref{fig5} (b), the  edge state entanglement spectrum has a cubic dispersion $E_k\propto k^3$ in the long wavelength limit, rather than linear. Therefore even in the long wavelength limit the entanglement Hamiltonian $H_E$ is {\it not} a chiral conformal field theory. It is reasonable to believe that the physical edge (left edge) is unaffected by such a change of dispersion, since it has zero entanglement temperature and stays in the ground state of the topological sector. However the entangled edge (right edge) is at finite entanglement temperature, so that its contribution to momentum should be modified when the dispersion is changed. To our surprise, the numerical results shown in Fig. \ref{fig5} (c) (d) clearly shows that formula (\ref{result}) still applies to this case. Such a robustness beyond edge CFT description indicates that the quantization of momentum polarization has a deeper origin from bulk topology.

\section{Projected wavefunctions for fractional Chern insulators}\label{sec:fci}

In this section we will calculate the topological spin and chiral
central charge for a $\nu =1/2$ lattice Laughlin state
\cite{Kalmeyer1987,Schroeter2007,Nielsen2012}. These lattice
Laughlin states, also known as the fractional Chern insulators, are
obtainable from parton constructions\cite{wen1991b,wen1999}, and we
develop a Monte Carlo approach for calculating their momentum
polarization. Based on Eq. (\ref{result}), our numerical results for
topological spin and chiral central charge are obtained for the $\nu
=1/2$ lattice Laughlin state, which agree very well with the
theoretical predictions from CFT. This suggests that momentum
polarization is a very useful quantity for identifying chiral
topological order from ground-state wave functions of interacting
systems.

The fractional Chern insulators are lattice analogies of fractional
quantum Hall states \cite{tang2011,sun2011,neupert2011,sheng2011}.
When bosons or fermions fractionally fill a nearly flat band with
nonzero Chern number, incompressible quantum liquid states with
fractional quantized Hall conductance can appear as stable ground
states of a repulsive interaction Hamiltonian. In this
work, we focus on a particular example of fractional Chern
insulator, i.e. a $\nu =1/2$ lattice Laughlin state of hardcore
bosons. Let us consider a $N\times N$ square lattice with single
parton sites. To construct a $\nu =1/2$ lattice Laughlin state, we
use the parton construction in Ref. \cite{zhang2011,zhang2012a} and
split the hardcore boson at each site into two fermionic partons,
$b_{i}^{\dagger}=c_{i\uparrow }^{\dagger }c_{i\downarrow}^{\dagger
}$. The physical Hilbert space of hardcore bosons requires either
two partons or no parton at each site, corresponding to the presence
or absence of hardcore boson, respectively. To build a hardcore
boson wave function from partons, a projector $P_{\mathrm{G}}$ is
needed to remove those unphysical configurations with single-parton
sites. Then, a hardcore boson state can be obtained by acting the
projector $P_{\mathrm{G}}$ on the parton wave function
$\left|\Phi_a\right\rangle = P_G \left|\phi_a \right\rangle_\uparrow
\left|\phi_a\right\rangle_\downarrow $.

\begin{figure}[tbp]
\centering
\includegraphics[width=2in]{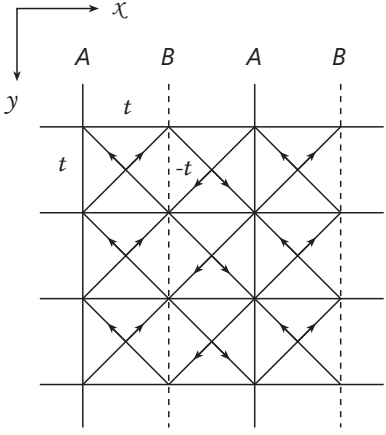}
\caption{Illustration of the Chern insulator Hamiltonian
(\ref{Chern}). The two sublattices in the unit cell are labeled by A
and B. The nearest-neighbor hopping integrals are $t$ along $x$
direction and odd columns along $y$ direction, and $-t$ along even
columns (dashed lines) along $y$ direction. The
next-nearest-neighbor hopping integrals are $\Delta $ ($-\Delta $)
if the hopping direction is along (against) the arrow.} \label{fig6}
\end{figure}

Now we assume that the fermionic partons are in the ground state of the following Chern insulator (See Fig. \ref{fig6})
\be H_C=\sum_{\langle ij\rangle ,\sigma }t_{ij}c_{i\sigma }^{\dagger
}c_{j\sigma }+i\sum_{\langle \langle ik\rangle \rangle ,\sigma
}\Delta _{ik}c_{i\sigma }^{\dagger }c_{k\sigma }  \label{Chern} \ee
Here the nearest-neighbor hopping integrals $t_{ij}$ are equal to
$t$ along $x$ direction and take the value $t$ ($-t$) in odd (even)
columns along $y$ direction. The next-nearest-neighbor hopping
integrals $\Delta _{ij}$ are $\Delta $ ($-\Delta $) if the hopping
direction is along (against) the arrow in Fig. \ref{fig6}. In this
work we focus on $t=1$ and $\Delta=1/2$. For periodic boundary
conditions along both directions, the Hamiltonian (\ref{Chern}) has
two bands with Chern number $C=\pm 1$. At half filling, the ground
state of partons has a completely filled lower band. After projecting
this parton band insulator state onto hardcore bosons, it has been shown
\cite{zhang2011} that the resulting projected wave function is a
$\nu =1/2$ lattice Laughlin state with two-fold degenerate on a
torus, and its topological sectors are related with the boundary
conditions of partons: the identity sector (semion sector) is given
by $|\Phi_1>=|0,0>+|\pi,0>=|0,\pi>+|\pi,\pi>$
($|\Phi_s>=|0,0>-|\pi,0>=|0,\pi>-|\pi,\pi>$), where
$|\Psi_x,\Psi_y>$ is the projected wavefunction with periodic
($\Psi_x=0$) or anti-period ($\Psi_x=\pi$) parton boundary
conditions along $ x$ and similarly $\Psi_y=0,\pi$ along $ y$.

The parton construction of the $\nu=1/2$ lattice Laughlin ground
states on a cylinder has further complication due to its gapless
chiral edge modes. To shed some light on the two-fold degenerate
ground-state projected wavefunctions, imagine that we start on a
torus and adiabatically lower all hopping amplitudes across the
$ x$ boundary until they are much smaller than the chiral edge
modes' finite size gap proportional to $L^{-1}$. As such a process
involves no gap closing or level crossing, the wavefunction's
topological properties, especially the eigenvalue of Wilson loops
along $y$ identifying the topological sectors should not
change. Consider the special case when the chiral edge modes on the
two edges of the cylinder have one exactly zero energy state each:
$c_{\sigma L}$ and $c_{\sigma R}$. On comparison with the open ${x}$ boundary
condition case, the small inter-edge hopping amplitudes will not
modify the chiral edge modes, except for the coupling of the two
zero energy states $c_{L}$ and $c_{R}$ in the form of
$\delta\left(\sum_\sigma c_{\sigma L}^{\dagger}c_{\sigma R}+h.c.\right)$. Here $\delta>0$ ( $\delta<0$) is remnant
from periodic (anti-periodic) boundary condition along $x$, respectively. In this way, the
degeneracy between $c_{L}$ and $c_{R}$ is lifted, and the
corresponding projected wavefunctions by filling all negative energy
parton states are
$\ket{0,\Psi_y}=P_G\left(c_{L\uparrow}^{\dagger}-c_{R\uparrow}^{\dagger}\right)\left(c_{L\downarrow}^{\dagger}-c_{R\downarrow}^{\dagger}\right)\left|\Phi'\right>$
and
$\ket{\pi,\Psi_y}=P_G\left(c_{L\uparrow}^{\dagger}+c_{R\uparrow}^{\dagger}\right)\left(c_{L\downarrow}^{\dagger}+c_{R\downarrow}^{\dagger}\right)\left|\Phi'\right>$,
respectively, where
$\left|\Phi'\right>=\prod_{\epsilon_\alpha<0}c^\dagger_\uparrow(\alpha)c^\dagger_\downarrow(\alpha)|
0 \rangle$ is obtained with all other parton states deeper in the
valence bands filled. As previous knowledge about the torus geometry
case would imply, the two topological sectors on a cylinder
correspond the symmetrization and anti-symmetrization of the
$\delta>0$ and $\delta<0$ cases, which give: \bea
|\Phi_1>&=&P_G\left(c_{L\uparrow}^{\dagger}c_{L\downarrow}^{\dagger}+c_{R\uparrow}^{\dagger}c_{R\downarrow}^{\dagger}\right)\left|\Phi'\right>=P_G c_{L\uparrow}^{\dagger}c_{L\downarrow}^{\dagger}\left|\Phi'\right>\\
|\Phi_s>&=&P_G\left(c_{R\uparrow}^{\dagger}c_{L\downarrow}^{\dagger}+c_{L\uparrow}^{\dagger}c_{R\downarrow}^{\dagger}\right)\left|\Phi'\right>=P_G
c_{L\uparrow}^{\dagger}c_{R\downarrow}^{\dagger}\left|\Phi'\right>\eea
In the last step of both equations we have used the particle hole symmetry of the Hamiltonian (\ref{Chern}). $\ket{\Phi_a},~a=1,s$ can be written in the equivalent  form of
$\left|\Phi_a\right\rangle = P_G \left|\phi_a \right\rangle_\uparrow
\left|\phi_a\right\rangle_\downarrow $, with the two parton
wavefunctions $\ket{\phi_a}_{\uparrow,\downarrow}$ defined by 
\bea
\left|\phi_1\right\rangle_\sigma&=& c^\dagger_{L\sigma}\prod_{\epsilon_\alpha<0}c^\dagger_\sigma(\alpha)| 0 \rangle\label{partonwf1}\\
\left|\phi_s\right\rangle_\sigma&=&
\frac{c^\dagger_{L\sigma}+ic^\dagger_{R\sigma}}{\sqrt
2}\prod_{\epsilon_\alpha<0}c^\dagger_\sigma(\alpha)| 0 \rangle
\label{partonwf2} \eea with $\sigma = \uparrow, \downarrow$.

The identification of topological sectors above is further confirmed by using
 a Monte Carlo algorithm\cite{zhang2012b} to calculate the
overlap between a series of projected wavefunctions of the form $\left|\Phi_{(p,q)}\right\rangle = P_G \left|\phi_{(p,q)} \right\rangle_\uparrow
\left|\phi_{(p,q)}\right\rangle_\downarrow $, with
\be
\left|\phi_{(p,q)}\right\rangle_\sigma=\left(p c^\dagger_{L\sigma}+
q
c^\dagger_{R\sigma}\right)\prod_{\epsilon_\alpha<0}c^\dagger_\sigma(\alpha)|
0 \rangle\ee
and $p,q$ complex coefficients. The overlap of the states with trial values $(p,q)=(0,1),~(1,0),~(1/\sqrt{2},\pm 1/\sqrt{2}),~(1/\sqrt{2},i/\sqrt{2})$ confirms that the only two linearly independent basis states are the two topological ground states $(p,q)=(1,0)$ and $(p,q)=(1/\sqrt{2},i/\sqrt{2})$, corresponding to Eq.
(\ref{partonwf1}) and (\ref{partonwf2}).

\begin{figure}[tbp]
\centering
\includegraphics[scale=0.4]{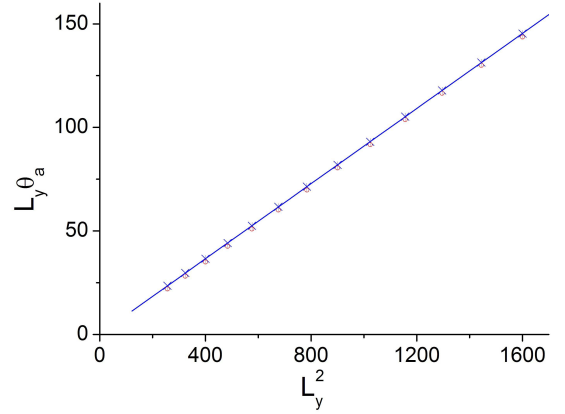}
\includegraphics[scale=0.4]{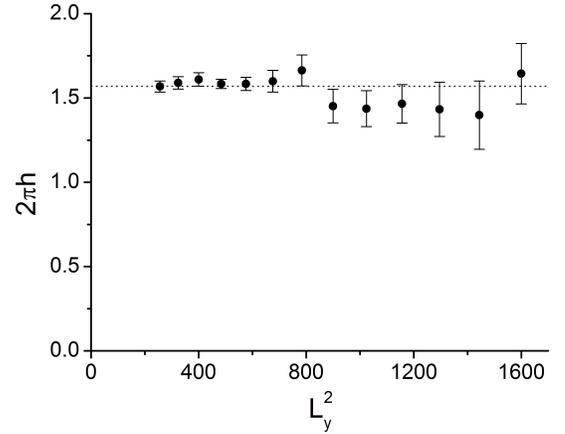}
\caption{Upper: The $L_y \theta_a$ data of various $L_y^2$ values
for the identity sector (blue crosses) and the quasiparticle sector
(red dots) of a fractional Chern insulator. The intercept of the
linear fits give the value of $2\pi(c/24-h_a)$. For the identity
sector, the data is well consistent with linear fit (blue solid
line) and suggests $c=1.078\pm0.091$; lower: the difference in $L_y
\theta_a$ corresponds to the measured values of $h_{s}=(L_y
\theta_1-L_y \theta_{s})/2\pi=0.2521\pm0.0063$, the dotted line is
the theoretical value $h_{s}=1/4$.} \label{fig:dataChern}
\end{figure}

To develop a Monte Carlo method to compute the momentum
polarization, we note that the expectation value of the twist
operator $T_{y}^{L}$ for $|\Phi_a \rangle $ can be express as \be
\langle \Phi_a |T_{y}^{L}|\Phi_a \rangle =\sum_{\alpha }|\langle
\alpha |\Phi_a \rangle |^{2}\frac{\langle \Phi_a |T_{y}^{L}\alpha
\rangle }{\langle \Phi_a |\alpha \rangle } \label{MCformula} \ee
where $|\alpha \rangle $ corresponds to a lattice configuration with
$N^{2}/2$ hardcore bosons, and $|T_{y}^{L}\alpha \rangle $ a
configuration with the positions of hardcore bosons at the left half
of the cylinder being translated by one lattice spacing in $ y$.
Based on Eq. (\ref{MCformula}), $\langle \Phi_a |T_{y}^{L}|\Phi_a
\rangle $ can be straightforwardly evaluated by treating $|\langle
\alpha |\Phi_a \rangle |^{2}$ as the probability for Monte Carlo
updates and $\langle \Phi_a |T_{y}^{L}\alpha \rangle /\langle
\Phi_a|\alpha \rangle $ as the Monte Carlo measurable. For the
present $\nu =1/2$ Laughlin state wavefunctions, these quantities
can be expressed in terms of determinant squares and updated
efficiently with the inverse-matrix update
techniques\cite{Ceperley1977}. We emphasize that this Monte Carlo
approach is also applicable for other chiral topological states from
fermionic parton constructions\cite{wen1989b}, whose wavefunctions
take the form of determinants and/or Pfaffians.

The results from this Monte Carlo method are shown in Fig.
\ref{fig:dataChern}. According to Eq. \ref{result}, the intercept of
the linear fit of $L_y \theta_a$ versus $L^2_y$ for the identity
sector corresponds to the measured central charge of the system
$c=1.078\pm0.091$, whereas the difference in $L_y \theta_a$ between
the two sectors corresponds to $2\pi h$, so the measured topological
spin is $h_{s}=0.252\pm0.006$. These are well consistent with the
known theoretical values for the model $c=1$ and $h=1/4$, and
suggest Eq. (\ref{result}) is also applicable to the parton
constructed systems.

\section{Discussion and conclusion}\label{sec:conclusion}

In summary, we have proposed a new efficient method for computing the quasi-particle topological spin and chiral central charge for chiral topologically ordered states. By defining the partial translation operator $T_y^L$ which translates the left part of the cylinder by one lattice constant, the combination of topological spin $h_a$ and chiral central charge $c$ can be extracted from a universal term $\frac{2\pi}{L_y}\left(h_a-\frac{c}{24}\right)$ in the phase of the ground state average values of $T_y^L$ in each topological sector. $h_a-\frac{c}{24}$ is the momentum polarization of the ground state in the topological sector $a$. We have derived the formula of central charge and topological spin (Eq. (\ref{result})) from the edge CFT form of the reduced density matrix, and also numerically verified the formula in two distinct topological states, the honeycomb lattice Kitaev model and the $\nu=1/2$ bosonic Laughlin state in fractional Chern insulators. Compared to previous numerical approaches which computes braid matrix of quasi-particles\cite{lahtinen2009,bolukbasi2012} and modular $S,T$ matrices\cite{zhang2012a,zhang2012b,cincio2012}, the method we propose is more efficient.

It is interesting to note that the momentum polarization defined on cylinder may be related to the characterization of one-dimensional symmetry protected topological states. If we view the cylinder as a one-dimensional (1D) chain, with the cylinder direction as an ``extra-dimension", the system can be viewed as a one-dimensional system with a $Z_N$ discrete symmetry. Here $Z_N$ is the translation along $y$ direction with $N=L_y$. In 1D there is no intrinsic topological state but there are symmetry-protected topological states, which have nontrivial edge states at each end of the 1D chain protected by global symmetries of the system. The type of topological state for a given symmetry group $G$ is classified by the inequivalent projective representations of $G$, or equivalently by the second cohomology of the group $H^2(G,U(1))$\cite{fidkowski2011b,turner2011,chen2011}. When the edge state carries a nontrivial projective representation of the group $G$, there will be a nontrivial series of group operations $g_i$ such that $g_Ng_{N-1}...g_1=1$ but the sequential action of $g_i$ on the edge state leads to a nontrivial phase factor $e^{i\theta}$. This phase factor will be canceled the other edge so that the whole system is still invariant in $G$. Compared with the momentum polarization we see that the idea is very similar. A phase $\theta=2\pi \left(h-\frac{c}{24}\right)$ is obtained by the left edge upon a $2\pi$ rotation of the cylinder, which is canceled by the right edge. However, the group cohomology of $Z_N$ (and its continuum limit $U(1)$) is trivial.\cite{chen2011b} Therefore if only the translation $Z_N$ symmetry is concerned, the projective representation labeled by the fractional momentum $h-\frac{c}{24}$ can be linearized, and no nontrivial classification is obtained. More information is required besides translation symmetry to give a fundamental reason for the accurate quantization of $h_a,~c$ in lattice models that is observed in the current work. Physically, the missing information is probably related to the locality of the model in 2D, which distinguishes a 2D model on the cylinder from a more generic 1D model with $Z_N$ symmetry.

In the two example systems studied in this paper, the ground states in each topological sector is obtained from the knowledge to these states. However, it should be noted that in general our method does not require knowledge on topological sectors. If $\left|G_\alpha\right\rangle,~\alpha=1,2,...,n$ denote the $n$ orthogonal ground states on the cylinder in an unknown basis, $\left|G_\alpha\right\rangle$ may be superpositions of different topological sectors $\left|G_a\right\rangle$. An important observation is that the partial translation $T_y^L$ does not change the topological sector, so that $\left\langle G_a\right|T_y^L\left|G_b\right\rangle\propto \delta_{ab}$ is diagonal. Physically, the reason is that topological sector $a$ cannot be measured by local probes but can only be measured by creating a pair of quasi-particles, taking one of them around the cylinder and then reannihilate them. We can choose the path of the quasi-particle so that it's far from the boundaries of the cylinder and the partition in the middle. As long as this is satisfied, such a quasi-particle motion is unaffected by the operation of partial translation $T_y^L$. Therefore the topological sector has to remain the same after partial translation. Consequently, in a generic basis one just needs to calculate the matrix $\Lambda_{\alpha\beta}=\left\langle G_\alpha\right|T_y^L\left|G_\beta\right\rangle$ and diagonalize it. The eigenvalues are $\lambda_a$ with the form of Eq. (\ref{result}).

Besides the examples given in the current work, the momentum
polarization can be applied a wide range of other topological
states. The Monte Carlo method of computing the momentum
polarization discussed in Sec. \ref{sec:fci} generally applies to
other projected wavefunctions obtained by various parton
constructions. In particular, this approach can be used to detect
topological spin and chiral central charge in non-Abelian chiral
topological states, which will be helpful in identifying the
topological order in more complicated many-body wave functions. For
instance, natural candidates for testing the momentum polarization
scheme include SU($n$)$_{k}$ states \cite{zhang2012b} and
SO($2n+1$)$_{1}$ states \cite{tu2012}, whose trial wavefunctions and
chiral edge CFTs are known. The momentum polarization results of
these non-Abelian states will be reported in future works.

Our approach may also have interesting applications in topological states described by density matrix renormalization group (DMRG)\cite{white1992}. DMRG applies to all gapped 1D states well, which is equivalent to finding a variational ground state with the form of matrix product states (MPS)\cite{rommer1997}. MPS refers to states of the form $\ket{\Psi}=\sum_{\left\{\nu_i\right\}}\left[T_1^{\nu_1}T_2^{\nu_2}...T_N^{\nu_N}\right]_{\alpha\beta}\ket{\left\{\nu_i\right\}}$,
with $\nu_i$ labels the physical states at each site, and $T^i_{\nu_i}$ a matrix with ``internal" indices $\alpha,\beta$. The DMRG/MPS approach has been applied to 2D topological states on a cylinder\cite{shibata2001,feiguin2008,jiang2012,stoudenmire2012}, which can be viewed as a 1D gapped state with a large number of states at each site, or one with a finite but long range interaction. Recently, the infinite DMRG, or infinite MPS, description has been studied for FQH states\cite{zaletel2012b} and FCI states\cite{cincio2012}, which simplifies the variational state by assuming a translation invariant ansatz and taking the infinite length limit. Besides the MPS states obtained numerically, the model wavefunctions of FQH states obtained from CFT correlation functions have also been shown to have the infinite MPS form when defined on a cylinder\cite{zaletel2012a,estienne2012} (with infinite internal dimension). The infinite MPS is particularly suitable for the study of the momentum polarization we propose, since it provides a description of the reduced density matrix using the internal auxiliary space (called the Schmidt states). In the long cylinder limit, the momentum polarization can be directly computed from the matrices $T^\mu$ at each site. An interesting open question is whether it is possible to obtain a generic proof of the universal quantization of the momentum polarization $h_a-\frac{c}{24}$ from the infinite MPS approach, which, if exists, will be more rigorous than the CFT approach taken in the current work. Another interesting direction is to generalize the infinite MPS approach of Ref. \onlinecite{zaletel2012b} to cylinder FQH to FCI states by using the Wannier state representation\cite{qi2011,wu2012c,lee2012}.

We would like to note that Ref. \onlinecite{cincio2012} has studied the modular $S,T$ matrices by a similar approach as Ref. \onlinecite{zhang2012a}, and Ref. \onlinecite{zaletel2012b} has independently studied the topological spin in an approach similar to ours, but with a continuous instead of discrete twist operator which is specially defined for cylinder FQH states.

{\bf Acknowledgement.} We would like to acknowledge the helpful discussions with Meng Cheng, J. Ignacio Cirac, Taylor L. Hughes, Anne E. B. Nielsen, Frank Pollmann, German Sierra, Zhenghan Wang and Edward Witten. This work is supported in parts by the EU project AQUTE (HHT), the Stanford Institute for Theoretical Physics (YZ) and the National Science Foundation through the grant No. DMR-1151786 (XLQ).

\bibliography{TI}

\begin{widetext}

\appendix

\section{Reduced density matrix and momentum polarization of Kitaev model}\label{app:kitaev}

We first review the structure of the reduced density matrix of Kitaev model obtained in Ref. \cite{yao2010}. Without losing generality, in the following we assume $L_y$ to be even. The first step is to separate the degrees of freedom to the left and right region. In the enlarged Hilbert space, the Majorana fermions $\eta_i$ and most link variables $u_{ij}$ are already separated into the two regions $L$ and $R$, and the only degrees of freedom that need special treating is the link variables $u_{ij}$ on the links crossing the boundary. By pairing up the neighbor links across the boundary, we can redefine the link variables as is shown in Fig. \ref{fig:app1} In the 4 boundary sites $1,2,3,4$, two link variables are defined as $u_{12}=i\gamma_1^z\gamma_2^z,~u_{34}=i\gamma_3^z\gamma_4^z$. Now we define the new link variables $w_{13}=i\gamma_1^z\gamma_3^z,~w_{42}=i\gamma_4^z\gamma_2^z$. It is easy to see that $w_{ij}$'s commute with each other and they both anticommute with $u_{ij}$'s. Also $w_{13}w_{42}=u_{12}u_{34}$. In the 4-dimensional Hilbert space of the 4 Majorana fermions (before projection), choosing the eigenstates of $u_{12},u_{34}$ we can find a basis in which $u_{12}=\sigma^z\otimes 1,~u_{34}=1\otimes \sigma^z$, $w_{13}=\sigma^x\otimes\sigma^x$, $w_{42}=-\sigma^y\otimes \sigma^y$. Therefore the eigenstates of $w_{13},w_{42}$ are
\bea
\ket{+-}&=&\frac1{\sqrt{2}}\left(\ket{\uparrow\downarrow}+\ket{\downarrow\uparrow}\right), ~\ket{++}=\frac1{\sqrt{2}}\left(\ket{\uparrow\uparrow}+\ket{\downarrow\downarrow}\right)\nonumber\\
\ket{--}&=&\frac1{\sqrt{2}}\left(\ket{\uparrow\uparrow}-\ket{\downarrow\downarrow}\right),~\ket{-+}=\frac1{\sqrt{2}}\left(\ket{\uparrow\downarrow}-\ket{\downarrow\uparrow}\right)\\
\ket{\uparrow\uparrow}&=&\frac1{\sqrt{2}}\left(\ket{++}+\ket{--}\right)
\eea
with the two $\pm$ labeling the eigenvalues of $w_{13},w_{24}$ respectively, and $\uparrow, \downarrow$ labels those of $u_{12},u_{34}$. A state $\ket{\left\{u_{ij}\right\}}$ with fixed values of $u_{ij}$ on all links can be written in the new basis. For simplicity we can always make a gauge choice so that $u_{ij}=1$ for all links crossing the boundary. For such $\left\{u_{ij}\right\}$, we have
\bea
\ket{\left\{u_{ij}\right\}}=2^{-L_y/4}\sum_{w_{ij}=\pm 1}\ket{\left\{u_{ij}^L,w_{ij}\right\}}_L\otimes \ket{\left\{u_{ij}^R,w_{ij}\right\}}_R
\eea
A physical ground state has the form
\bea
\ket{G}=P\left(\ket{G_F\left(\left\{u_{ij}\right\}\right)}\otimes \ket{\left\{u_{ij}\right\}}\right)
\eea
with $P=\prod_i\frac{D_i+1}{2}$. We write the $\eta_i$ fermion ground state $\ket{G_F\left(\left\{u_{ij}\right\}\right)}$ in the Schmit decomposition form
\bea
\ket{G_F\left(\left\{u_{ij}\right\}\right)}=\sum_N\alpha_N\ket{\Psi_N^L\left(\left\{u_{ij}^L\right\}\right)}\ket{\Psi_N^R\left(\left\{u_{ij}^R\right\}\right)}
\eea
with $\ket{\Psi^L_N}$ and $\ket{\Psi_N^R}$ a set of orthogonal basis of left and right side $\eta$ fermions respectively, and $\alpha_N$ the Schmit eigenvalues satisfying $\sum_N\abs{\alpha_N}^2=1$. Such a Schmit decomposition for quadratic fermion problem can be reduced to a simple single-particle problem\cite{peschel2003}, the detail of which will be discussed later in this appendix. Thus the ground state is written as
\bea
\ket{G}=2^{-L_y/4}\sum_{w_{ij}=\pm 1}\sum_N\alpha_NP_L\left[\ket{\Psi_N^L\left(\left\{u_{ij}^L\right\}\right)}\ket{\left\{u_{ij}^L,w_{ij}\right\}}_L\right]\otimes P_R\left[\ket{\Psi_N^R\left(\left\{u_{ij}^R\right\}\right)}\ket{\left\{u_{ij}^R,w_{ij}\right\}}_R\right]
\eea
in which we have written $P=P_LP_R$ since the projection is local and separable to two subsystems. If we expand $P_R=\prod_{i\in R}\frac{D_i+1}2$, there are two terms which preserve the gauge field configurations $u_{ij}^R, w_{ij}$, which are $1$ and $\prod_{i\in R}D_i$. The second term does a gauge transformation to all sites $\eta_i\rightarrow -\eta_i$ but preserves all the link variables. Therefore we can write
\bea
P_R\left[\ket{\Psi_N^R\left(\left\{u_{ij}^R\right\}\right)}\ket{\left\{u_{ij}^R,w_{ij}\right\}}_R\right]&=&\sum_{\left\{{u'}_{ij}^{R},w_{ij}'\right\}\simeq \left\{{u}_{ij}^{R},w_{ij}\right\}}\frac{1+\prod_{i\in R}D_i}2\ket{\Psi_N^R\left(\left\{{u'}_{ij}^R\right\}\right)}\ket{\left\{{u'}_{ij}^R,w'_{ij}\right\}}_R\label{projected}
\eea
and similar for $P_L$. Here $\simeq$ denotes gauge equivalence. For two states $P_R\left[\ket{\Psi_N^R\left(\left\{u_{ij}^R\right\}\right)}\ket{\left\{u_{ij}^R,w_{ij}\right\}}_R\right]$ with different boundary configurations $w_{ij}$, after the projection they are still orthogonal since it is not possible to do a gauge transformation to $w_{ij}$ without affecting $u_{ij}$ on neighboring links. Also the number of orthogonal states in Eq. (\ref{projected}) is the same for all configurations. Therefore the reduced density matrix is
\bea
\rho_L&=&Const.\cdot \sum_{w_{ij}=\pm 1}\sum_N\abs{\alpha_N}^2P_L\left[\ket{\Psi_N^L\left(\left\{u_{ij}^L\right\}\right)}\ket{\left\{u_{ij}^L,w_{ij}\right\}}_L\right]\left[\bra{\Psi_N^L\left(\left\{u_{ij}^L\right\}\right)}\bra{\left\{u_{ij}^L,w_{ij}\right\}}_L\right]P_L\nonumber\\
&=&Const.\cdot \sum_{\left\{{u'}_{ij}^{L},w_{ij}'\right\},~\left\{{u''}_{ij}^{L},w_{ij}''\right\}\simeq \left\{{u}_{ij}^{L},w_{ij}\right\}}\sum_N\abs{\alpha_N}^2
\frac{1+\prod_{i\in L}D_i}2\ket{\Psi_N^L\left(\left\{{u'}_{ij}^L\right\}\right)}\ket{\left\{{u'}_{ij}^L,w'_{ij}\right\}}_L
\bra{\Psi_N^L\left(\left\{{u''}_{ij}^L\right\}\right)}\bra{\left\{{u'}_{ij}^L,w''_{ij}\right\}}_L\frac{1+\prod_{i\in L}D_i}2
\eea
The translation operator $T_y^L$ translates both $\eta_i$ and $u_{ij},w_{ij}$. The global projection $\frac{1+\prod_{i\in R}D_i}2$ only cares about the total fermion number parity, and commutes with $T_y^L$. Thus
\bea
\lambda&\equiv &{\rm Tr}\left(T_y^L\rho_L\right)=Const.\cdot \sum_{w_{ij}=\pm 1}\sum_{\left\{{u'}_{ij}^{L},w_{ij}'\right\}\simeq \left\{{u}_{ij}^{L},w_{ij}\right\}}\sum_N\abs{\alpha_N}^2
\bra{\Psi_N^L\left(\left\{T{u'}_{ij}^L\right\}\right)}\frac{1+\prod_{i\in L}D_i}2T_y^{L\eta}\ket{\Psi_N^L\left(\left\{{u'}_{ij}^L\right\}\right)}\label{kitaevmompol1}
\eea
Here $T_y^{L\eta}$ denotes the action of $T_y^L$ to $\eta$ subspace, and $T{u'}_{ij}^L$ is configuration obtained from the translation of ${u'}_{ij}^L$ by one lattice constant. For all gauge equivalent configurations $\left\{{u'}_{ij}^L\right\}$, the matrix element in Eq. (\ref{kitaevmompol1}) has the same value, since the gauge transformations to ${u'}$ and $T{u'}$ cancels each other after being acted by $T_y^L$. Therefore we can just take the representative configuration $u_{ij}$ and write
\bea
\lambda&\equiv &{\rm Tr}\left(T_y^L\rho_L\right)=Const.\cdot \sum_{w_{ij}=\pm 1}\sum_N\abs{\alpha_N}^2
\bra{\Psi_N^L\left(\left\{T{u}_{ij}^L\right\}\right)}\frac{1+\prod_{i\in L}D_i}2T_y^{L\eta}\ket{\Psi_N^L\left(\left\{{u}_{ij}^L\right\}\right)}
\eea
Now we do some more analysis to $D_i$. We can write
\bea
\prod_{i\in L}D_i=F_\eta \prod_{\avg{ij}\in L}u_{ij}\prod_{ij}w_{ij}
\eea
with $F_\eta$ the fermion number parity of $\eta$ fermions. To write down this formula some suitable orientation convention should be made to the links since $u_{ij}=-u_{ji}$, but the detail of this convention is not important here. The projector $\frac{1+\prod_{i\in L}D_i}2$ imposed the constraint $\prod_{i\in L}D_i=1$ which requires $F_\eta =\prod_{\avg{ij}\in L}u_{ij}\prod_{ij}w_{ij}$. Therefore in the sum over $2^{L_y/2}$ possible configurations of $w_{ij}$, half of them has the fermion projected to $F_\eta=1$ and the other half projected to $F_\eta=-1$. Since the Schmit state $\ket{\Psi_N^L\left(\left\{{u}_{ij}^L\right\}\right)}$ is independent from the boundary variable $w_{ij}$, the sum over the two projections simply give
\bea
\lambda=\sum_N\abs{\alpha_N}^2
\bra{\Psi_N^L\left(\left\{T{u}_{ij}^L\right\}\right)}T_y^{L\eta}\ket{\Psi_N^L\left(\left\{{u}_{ij}^L\right\}\right)}
\eea
Defining the gauge transformation $V\left(\left\{{u}_{ij}^L\right\}\right)$ by
\bea
V\left(\left\{T{u}_{ij}^L\right\},\left\{{u}_{ij}^L\right\}\right)\ket{\Psi_N^L\left(\left\{{u}_{ij}^L\right\}\right)}=\ket{\Psi_N^L\left(\left\{T{u}_{ij}^L\right\}\right)}
\eea
and the covariant translation operator
\bea
T_y^{LF}=V^\dagger\left(\left\{T{u}_{ij}^L\right\},\left\{{u}_{ij}^L\right\}\right)T_y^{L\eta}
\eea
we have
\bea
\lambda&=&{\rm Tr}\left(T_y^{LF}\rho_{LF}\right)\\
\text{with~}\rho_{LF}&=&\sum_N\abs{\alpha_N}^2\ket{\Psi_N^L\left(\left\{{u}_{ij}^L\right\}\right)}\bra{\Psi_N^L\left(\left\{{u}_{ij}^L\right\}\right)}
\eea

To see the effect of $T_y^{LF}$ more explicitly, we discuss the periodic and antiperiodic sectors separately. For the periodic sector with $\prod_{\avg{ij}}u_{ij}=1$ around the cylinder, we can simply take the gauge choice $u_{ij}=1$ for all links. Since this gauge is already translation invariant, $T_y^{LF}$ is the ordinary translation operator. The momentum eigenstates $f_{nk}$ has the wavefunction
\bea
f_{nk}=\sum_{i}\phi_{nk}(i_x)e^{-iky_i}\eta_i
\eea
with $k=\frac{2\pi}{L_y}n,~n\in\mathbb{Z}$, and $(i_x,i_y)$ the two-dimensional coordinate of $i$ site. The translation operator acts as
\bea
{T_y^{LF}}^{-1}\eta_iT_y^{LF}=\eta_{i+\hat{y}},~{T_y^{LF}}^{-1}f_{nk}T_y^{LF}=e^{ik}f_{nk}\label{TyPBC}
\eea

The anti-periodic sector is a bit more complicated. We can choose a gauge with $u_{ij}=-1$ on all links across a horizontal line, as is shown by the black solid line in Fig. \ref{fig:app1} (b), and $u_{ij}=1$ on all other links. For convenience we label the horizontal chains parallel to the branchcut line by $n=1,2,...,L_y$ as is illustrated in Fig. \ref{fig:app1} (b) by the blue solid line. We choose the labeling such that the branchcut line is at the boundary, between $n=1$ and $n=L_y$.  In this gauge, the gauge transformation $V\left(\left\{T{u}_{ij}^L\right\},\left\{{u}_{ij}^L\right\}\right)$ needs to translate the branchcut line by one lattice constant, which is defined by
\bea
V\eta_iV^\dagger =-\eta_i,~\forall~i\in \text{~chain~}1
\eea
Now if we define
\bea
f_{nk}=\sum_{i}\phi_{nk}(i_x)e^{-iky_i}\eta_i
\eea
with $y_i\in[1,L_y]$ and $k=\frac{2\pi}{L_y}\left(n-\frac12\right),~n=1,2,...,L_y$, the function $e^{-iky_i}$ has a jump between $y_i=1$ and $y_i=L_y$. Therefore
\bea
{T_y^{L\eta}}^{-1}f_{nk}T_y^{L\eta}&=&\sum_{i_x}\left[\sum_{i_y=2}^{L_y}\phi_{nk}(i_x)e^{-ik\left(y_i-1\right)}\eta_i+\phi_{nk}(i_x)e^{-ikL_y}\eta_{(i_x,1)}\right]\nonumber\\
&=&e^{ik}\sum_{i_x}\left[\sum_{i_y=2}^{L_y}\phi_{nk}(i_x)e^{-iky_i}\eta_i-\phi_{nk}(i_x)e^{-ik}\eta_{(i_x,1)}\right]
\eea
Thus we see that the gauge transformation $V$ correctly removes the additional minus sign and obtain
\bea
{T_y^{LF}}^{-1}f_{nk}T_y^{LF}=e^{ik}f_{nk}\label{TyABC}
\eea
Due to Eq. (\ref{TyPBC}) and (\ref{TyABC}) we see that $T_y^{LF}$ in the particular gauge choice can be written in the form of Eq. (\ref{Tyfree})
 \bea
 T_y^{LF}=\exp\left[i\sum_{n,k}kf_{nk}^\dagger f_{nk}\right]
 \eea
 for both periodic and anti-periodic flux sectors. Using the fact that $f_{nk}^\dagger f_{nk}$ has eigenvalues $0,1$ we can write
 \bea
 T_y^{LF}=\prod_{nk}\exp\left[ikf_{nk}^\dagger f_{nk}\right]=\prod_{nk}\left[f_{nk}^\dagger f_{nk}e^{ik}+\left(1-f_{nk}^\dagger f_{nk}\right)\right]
 \eea
 Using this equation and the entanglement Hamiltonian (\ref{HEfree}), it is straightforward to obtain Eq. (\ref{lambdafree}).

\begin{figure}[tbp]
\centering
\includegraphics[scale=0.4]{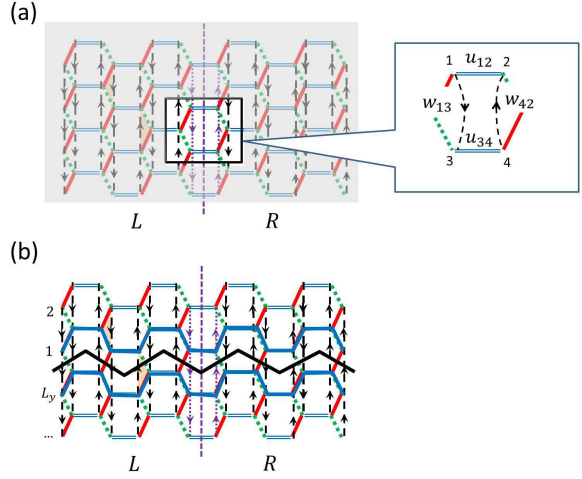}
\caption{(a) The definition of boundary variables $w_{ij}$ on four sites $1,2,3,4$. The boundary is drawn as the purple vertical dashed line. (b) Illustration of the gauge choice for anti-periodic boundary condition, with all $u_{ij}=1$ except those crossing the branchcut line (thick black solid line). The sites on the lattice are organized into chains parallel to the branchcut line, shown by the blue thick solid lines. The chain labeling is given on the left.  } \label{fig:app1}
\end{figure}

As the last part of this appendix we briefly review the free fermion reduced density matrix $\rho_{LF}$ calculated in Ref. \cite{peschel2003}. All multi-point functions $\bra{G}\prod_{i=1}^{2N}\eta_i\ket{G}$ of the free fermion ground state satisfies the Wick theorem, so that all correlation functions are determined by the two-point function
\bea
C_{ij}=\bra{G}\eta_i\eta_j\ket{G}-\delta_{ij}
 \eea
 In particular, $C_{ij}$ for $i,j\in L$ in the left half cylinder determines all multi-point correlation functions in the left half cylinder. Therefore the reduced density matrix must i) also satisfy Wick theorem, and ii) reproduces the two-point functions. The first condition requires $\rho_{LF}$ to have the quadratic (thermal) form
\bea \rho_{LF}=e^{-H_E}=\exp\left[-\frac12\sum_{i,j\in
L}h_{ij}\eta_i\eta_j\right] \eea If we diagonalize $H_E$ into the
form given in Eq. (\ref{HEfree}) \bea
H_E=\sum_{n,k}\xi_{nk}\left(f_{nk}^\dagger f_{nk}-\frac 12\right)
\eea the correlation function will be \bea {\rm
Tr}\left(\rho_{LF}f_{nk}^\dagger
f_{nk}\right)=\frac1{e^{\xi_{nk}}+1} \eea and corresponding ${\rm
Tr}\left(\rho_{LF}f_{nk} f_{nk}^\dagger\right)$, which should agree
with the eigenvalues of $C_{ij}$ with the same eigenvectors.
Therefore in the matrix level we have \bea C=\tanh h~\Rightarrow
h=\frac12\left[\log \left(\mathbb{I}+C\right)-\log
\left(\mathbb{I}-C\right)\right] \eea This equation determines $h$
and its spectrum $\xi_{nk}$ from correlation function $C$.

\end{widetext}
\end{document}